\documentclass[11pt,twoside]{article}
\usepackage{atmp}

\usepackage{amsfonts}
\usepackage{graphicx}
\usepackage{hhline}
\usepackage{cite}

\def\be{\begin{equation}}
\def\ee{\end{equation}}
\def\bdm{\begin{displaymath}}
\def\edm{\end{displaymath}}
\def\bea{\begin{eqnarray}}
\def\eea{\end{eqnarray}}

\def\SE{Schr\"odinger equation}
\def\SO{Schr\"odinger operator}

\newcommand{\dsx}{{\displaystyle x}}
\newcommand{\prt}{{\tt p}}

\newcommand{\rd}{\mbox{d}}

\newcommand{\re}{\mbox{e}}

\newcommand{\ftnref}[1]{\mbox{$^{\ref{#1}}$}}

\begin{document}

\copyrightnotice{2003}{7}{711}{725}

\title{Higher-level eigenvalues of
Q-operators and \SE}
\url{hep-th/0307108}
\setcounter{page}{711}
\author{V.V.~Bazhanov$^1$, S.L.~Lukyanov$^{2,3}$ and A.B.~Zamolodchikov$^{2,3}$ }
\address{$^1$Department of Theoretical Physics,\\
Research School of Physical Sciences and Engineering,\\
IAS, Australian National University,\\
Canberra, ACT 0200, Australia\\\ \\
$^2$NHETC, Department of Physics and Astronomy\\
     Rutgers University\\
     Piscataway, NJ 08855-0849, USA\\ \ \\
$^3$L.D. Landau Institute for Theoretical Physics\\
  Chernogolovka, 142432, Russia}

\markboth{\it Higher-level eigenvalues of
Q-operators and \SE}{\it V.V.~Bazhanov, S.L.~Lukyanov and A.B.~Zamolodchikov}

\begin{abstract} 
Relation between one-dimensional 
\SE\ and the vacuum eigenvalues of
the  ${\bf Q}$-operators 
is extended to their higher-level eigenvalues.
\end{abstract}

\section{Introduction}

Remarkable relation between the spectral 
theory of \SE\ \cite{vorosa}
and vacuum eigenvalues 
of the so-called ${\bf Q}$
operators of Conformal Field Theory (CFT) was discovered 
a few years ago in \cite{DTa}.
The ${\bf Q}$-operators were introduced in 
\cite{BLZb}\  as a mathematical
tool to describe
integrable (Yang-Baxter) structure 
of $c\leq1$ CFT, where they play the
role of
the Baxter's $Q$-matrix \cite{BXT}. 
It was observed in \cite{DTa}\  that for several
degenerate Virasoro module the vacuum eigenvalues of
the ${\bf Q}$-operators 
essentially coincide with the spectral
determinants of the \SO\ $-{{\rd^2}\over{\rd \dsx^2}} + V(x)$
with the potential $V(x)=x^{2\alpha}$, where the exponent $\alpha$ is
determined by the
Virasoro central charge $c$. This relation was
extended in \cite{BLZf}\ to all Virasoro modules with $c<1$.
On the other hand, the Virasoro vacuum is one among a sequence of
eigenstates of the operators ${\bf Q}$ in 
the whole Virasoro module. The
question arises if such relation to the 
\SO\ exists for the higher-level eigenvalues.

In this note we answer this question positively
and identify the Schr\"odinger operators 
associated with the higher-level eigenvalues. 
For an eigenstate
on
the level $L$ of the 
highest-weight Virasoro module the potential has the
form
\bea\label{Vpot}
V(x)={\ell(\ell+1)\over x^2}+x^{2\alpha}-2\ {\rd^2\over
\rd x^2}\ \sum_{k=1}^L\,  \log(x^{2\alpha+2}-z_k)\ ,
\eea
where $\alpha$ and $\ell$ 
are related to the central charge $c$ and the
highest weight $\Delta$ of the Virasoro module as follows
\bea\label{cdelta}
c = 1 -{6\, \alpha^2\over \alpha + 1}\ , 
\qquad
\ \ \Delta={{(2\ell +1)^2}-4\alpha^2\over {16\,(\alpha + 1)}}\ .
\eea
The $z_k,\ k=1,2,\ldots L$ are pairwise different 
{\it nonzero} complex
numbers satisfying the system of $L$ algebraic equations
\bea\label{algsys}
\sum_{{j=1}\atop{j\not= k}}^{L}
{z_k\big( z_k^2 +(3+\alpha)(1+2\alpha)z_k z_j+
\alpha (1+2 \alpha)
z_j^2 \big)\over (z_k-z_j)^3}-{ \alpha\, z_k\over 4 (1+\alpha)}+
\Delta=0\ ,
\eea
with
$k=1,2,\ldots L$.
In what follows we refer to the potentials \eqref{Vpot}\ as the
``Q-potentials''.
The arguments presented below strongly 
suggest that appropriately defined
spectral determinants of the 
\SO\ $-{\rd^2\over \rd \dsx^2} + V(x)$
with
the Q-potentials $V(x)$ 
coincide, up to certain trivial factors, with the
eigenvalues of the ${\bf Q}$-operators.

The\ \SE\ with the monstrous
potentials \eqref{Vpot}\ are unlikely to be of much direct interest in
quantum mechanics. On the
other hand, the higher-level eigenvalues of the ${\bf Q}$-operators
are definitely of interest in integrable quantum field theories.
One of the reasons is their relation\ \cite{BLZa,BLZb}
to the boundary states associated with certain integrable
boundary flows\ \cite{Affleck,Fendley}.
This work was motivated by possible applications
to this kind of problems.

We did not attempt at 
making this paper self-contained. The definitions of the 
${\bf Q}$-operators and the local and nonlocal integrals of
motion are as in Ref.\cite{BLZb}. The arguments
in Section 3 are  meant to be read in conjunction with Refs.\cite{DTa,BLZf}.

\section{The ${\bf Q}$-operators}

Here we only list some basic properties of the ${\bf Q}$-operators
used in the discussion below. We refer the reader to 
\cite{BLZb,BLZc}
for explicit construction of the ${\bf Q}$-operators and more details.

The ${\bf Q}$-operators are actually the operator-valued functions
${\bf Q}_{\pm}(s)$ of a complex parameter 
$s\, $\footnote{The operators
${\bf Q}_{\pm}(s)$ discussed here
coincide with ${\bf Q}_{\pm}(\lambda)$ of \cite{
BLZb,BLZc}
with  
$$\lambda^2=\bigg[\, \sqrt{\pi}\  {\Gamma({1+\alpha\over 2\alpha})\over 
\Gamma({1\over 2\alpha})}\,
\bigg]^{
2\alpha\over 1+\alpha}\ 
\Gamma^{-2}\Big({\alpha\over 1+\alpha}\Big)\ s\ .
$$
The
parameter $\alpha$ is related to $\beta$ used in \cite{
BLZb,BLZc} as
$\beta^2={1\over 1+\alpha}$.}.
The operators ${\bf Q}_{\pm}(s)$ act
in the Virasoro module ${\cal V}_\Delta$. It will be convenient to
parametrize its highest
weight $\Delta$ and the Virasoro central charge $c <1$ as
in Eqs.\eqref{cdelta},
where we assume that the real parameters $\ell$ and $\alpha$ 
are in the range
$\ell\geq -1/2$, $\alpha>0$. The operators ${\bf Q}_\pm(s)$ form
a commutative family  
and they also commute with the
zero-mode Virasoro generator ${\bf L}_0$. The Virasoro module ${\cal
V}_{\Delta}$ admits the standard  level decomposition,
\bea\label{leveldef}
{\cal V}_\Delta=\oplus_{L=0}^\infty\, {\cal V}_\Delta^{(L)}\, ,\qquad
\qquad
{\bf L}_0\, {\cal V}_\Delta^{(L)}=(\Delta+L)\ {\cal V}_\Delta^{(L)}\, ,
\eea
and the level subspaces ${\cal V}_{\Delta}^{(L)}$ are invariant
with respect to the action 
of  the ${\bf Q}$-operators, 
\bea
{\bf Q}_{\pm}\,:\ {\cal
V}_{\Delta}^{(L)} \to  {\cal V}_{\Delta}^{(L)}\, .
\eea
In particular, the
Virasoro vacuum $|\, \Delta\,  \rangle$ is an eigenvector of the ${\bf
Q}$-operators, 
\bea
{\bf Q}_{\pm}(s)\ |\,  \Delta\, \rangle = Q_{\pm}^{(vac)}
(s)\ |\, \Delta\,  \rangle\, .
\eea
It is the vacuum eigenvalues $Q_{\pm}^{(vac)}(s)$
that have received most of attention in the Refs.\cite{DTa,BLZf}.
Here we will
discuss generic eigenvalues $Q_{\pm}(s)$ of the operators ${\bf
Q}_{\pm}(s)$ in ${\cal V}_{\Delta}$. 
As the functions of $s$ they enjoy
the following basic properties\ \cite{BLZb,BLZc}
\footnote
{\label{kjsui} In writing Eqs.\eqref{aslead},\eqref{wronsb}\ below
we assume that ${(1+\alpha)/ (2\alpha)}$ is not
an integer. If it equals to $N=1,\, 2,\, \ldots$, the asymptotic
form \eqref{aslead}\ changes to $s^N\, \log(-s)$ and
\eqref{wronsb} must be modified accordingly.
The free-fermion point $\alpha=1$ (i.e. $c=-2$)
is one of these ``resonance''
cases.};

\begin{itemize}

\item[${(}i{)}$]
The functions ${A}_{\pm}(s) = (-s)^{\mp {2\ell+1\over  4}}
\,{Q}_{\pm}(s)$ are
entire functions of $s$, and 
\bea\label{normalki}
A_{\pm}(0) =1\ .
\eea

\item[${(}ii{)}$]
The eigenvalues ${Q}_{\pm}(s)$ have the leading asymptotic
behavior:
\bea\label{aslead}
\log {Q}_\pm(s)=
{\pi\over \cos({\pi\over 2\alpha})}\  (-s)^{1+\alpha\over 2\alpha}+O(1)
\eea
at\ $\ |s| \to\infty\,, \  {\rm arg} (-s) < \pi$.

\item[${(}iii{)}$]
The functions ${A}_{\pm}(s)$ satisfy the functional
relation (the ``quantum 
Wronskian condition'')
\bea\label{wronsb}
q^{\ell+{1\over 2}} A_+(qs) A_-(q^{-1}s )-
q^{-\ell-{1\over 2}} A_+(q^{-1}s ) A_-(qs)= {q^{\ell+{1\over 2}}-
q^{-\ell-{1\over 2}}}\ ,
\eea
where
\bea\label{qdef}
q=\re^{ { {\rm i}\, \pi\over 1+\alpha }}\, .
\eea
\end{itemize}

\noindent
The equation \eqref{wronsb}\ follows directly from the
construction of the ${\bf Q}$-operators \cite{BLZb,BLZc}.
The statements of
analyticity, $(i)$ and $(ii)$ above, 
and especially the domain of validity
of the asymptotic form \eqref{aslead},
presently have somewhat less solid status. For the vacuum eigenvalues
$Q_{\pm}^{(vac)}(s)$ these properties follow from the results of 
\cite{BLZf}.
The solution at $c=-2$, where all
eigenvalues can be found explicitly\ \cite{BLZb}, 
also supports the above
analyticity statement $(i),\, (ii)$ (see footnote\ftnref{kjsui} on 
this page). 
But it is fair to say that a solid
proof in a general case is still missing. Nonetheless, in the further
discussion we assume these properties.

Entire functions $A_{\pm}(s)$ having the asymptotic 
behavior \eqref{aslead}
are  characterized by positions of their respective zeros
$s_{k}^{\pm}$ in the complex $s$-plane. The zeros accumulate towards
$s=\infty$ along the positive real axis. The quantum Wronskian
equation \eqref{wronsb}\ then leads to infinite set of the Bethe Ansatz
equations for the numbers $s_{k}^{\pm}$. The analysis of these equations
in Appendix~A shows that there is a discrete (albeit infinite) set
of functions $Q_{\pm}(s)$ that satisfy all the 
conditions $(i-iii)$ above.
Clearly, all eigenvalues of the ${\bf Q}$-operators can be found among
this set. But there is much evidence that the converse is also true ---
any pair of functions $Q_{\pm}(s)$ satisfying $(i-iii)$
are eigenvalues
of the pair of operators ${\bf Q}_{\pm}(s)$ in ${\cal V}_{\Delta}$. 
Given a solution of
$(i-iii)$, 
the level $L$ of the associated eigenvector can be identified
as follows. Corrections to the leading asymptotics 
\eqref{aslead}\ are
expressed through the 
eigenvalues of the local and nonlocal integrals of
motion, according to the 
expansion \eqref{asympt}\  below. Since the first local
integral ${\bf I}_1$ coincides with the 
Virasoro generator ${\bf
L}_0-c/24$, the first subleading terms 
to the asymptotics \eqref{aslead}\ have the
form $(\alpha>1/2)$:
\bea\label{sublead}
\renewcommand{\arraystretch}{2.}
\log Q_{\pm} (s)  &\simeq&{\pi\over \cos({\pi\over 2\alpha})}\
 (-s)^{1+\alpha\over 2\alpha} + C\nonumber \\
&&\ -{1\over \sin({\pi\over 2\alpha})}\
 \left(\, \Delta-{{c\over 24}}+L\, \right)
 (-s)^{-{1+\alpha\over 2\alpha}} 
+\ldots\,,
\eea
where the constant term 
does not depend on $s$ (but, of course, depends on
a state from $V_{\Delta}$) 
and $L$ is the level. The results of Appendix~A suggest that
for generic $c$ and $\Delta$ the number of those solutions of 
$(i-iii)$
which also have the subleading 
behavior \eqref{sublead}\ with a given
integer
$L\ge0$ equals the number of integer partitions of $L$. It thus coincides
with the dimension of the level subspace ${\cal V}_{\Delta}^{(L)}$ of
a generic non-degenerate Virasoro module ${\cal V}_{\Delta}$.

\section{Schr\"odinger equation}

It is known\ (see e.g.\cite{vorosa,BLZf})\ 
that appropriately defined spectral
determinants $D_{\pm}(E)$ 
(Eq.\eqref{sdets}
below)
for the\ \SE
\bea\label{Seq}
-{\rd^2\over \rd x^2}\, \Psi(x)+V(x)\,\Psi(x)=E\,\Psi(x)
\eea
with the potential
\bea\label{vacuumv}
V(x) = V^{(vac)}(x) \equiv {{\ell(\ell+1)}\over{x^2}} + x^{2\alpha}
\eea
obey all the above  properties $(i-iii)$, if one 
makes the identifications
\bea\label{conj}
D_\pm(\nu s)=A_\pm(s)\, ,
\eea
where
\bea\label{nudef}
\nu =\bigg[\, {2\sqrt{\pi}\, \Gamma({3\over 2}+{1\over 2\alpha})\over
\Gamma(1+{1\over 2\alpha})}\,
\bigg]^{
2\alpha\over 1+\alpha}\ .
\eea
Thus $D_{\pm}(E)$  are related
to some eigenvalues of the ${\bf Q}$-operators.
In fact,  the spectral determinants of
\eqref{Seq} with $V(x)=V^{(vac)}(x)$ correspond
to the {\it vacuum} eigenvalues $Q_{\pm}^{(vac)}$.
This correspondence in the case
$\ell=0$ is the original observation of \cite{DTa}.
It was proven in\ \cite{BLZf} 
by combining
Eq.\eqref{wronsb} with known
analyticity of
$Q_{\pm}^{(vac)}$ as functions of the
parameter $\ell$.

It is important that the properties $(i-iii)$
of the spectral determinants can be  derived
using only
the following three 
properties of the
potential \eqref{vacuumv}\ as the function of complex $x$:

\begin{itemize}

\item{$(I)$}   Symmetry:
\bea\label{Vsym}
V(q x)=q^{-2}\, V(x)\ ,
\eea
where $q$ is defined in \eqref{qdef}.

\item{$(II)$}
Asymptotic behavior:
\bea\label{asbeh}
&&V(x)\to {\ell(\ell+1)\over x^2}+o(1) \qquad {\rm as} \quad x\to 0\ ,
\\
&&V(x)\to x^{2\alpha}+o(1)\qquad \qquad
 {\rm as}\quad x\to\infty\ .
\eea

\item{$(III)$}
Monodromy properties: For any value of $E$ all solutions
of  \eqref{Seq}\ with this potential are monodromy free
everywhere except for $x=0$ and $x=\infty$.

\end{itemize}

\noindent
Here we observe 
that $V^{(vac)}(x)$ is not the only potential which
meets all of the conditions $(I-III)$. 
We show in  Appendix B that
the most general potential $V(x)$ with these properties 
is given by
\eqref{Vpot}, with the parameters 
$z_k$ satisfying the Eqs.\eqref{algsys}.
Note that for
$L >0$ the potential 
\eqref{Vpot}\ has poles away from zero, at $x^{2\alpha+2}=z_1,\, z_2,\,
\ldots, z_L$. 
The equations \eqref{algsys}\ then guarantee 
that these poles do not
violate\ $(III)$. 
Thus $(I-III)$ are the characteristic properties of all
Q-potentials.

Given a Q-potential, Eqs.\eqref{Vpot},\eqref{algsys},
the associated
spectral determinants $D_{\pm}(E)$ are defined as follows.
For $\Re e\,  \ell>(1+\alpha)\, L-{3/2}$
the equation \eqref{Seq}\ has a solution $\psi(x,\ell,E)$ uniquely
specified by the condition
\bea\label{jsusdy}
\psi(x,E,\ell):\ \ \ \psi(x,E, \ell)=x^{\ell+1}+O(x^{\ell+3})\ \ \ \
{\rm as}\ \ \ \ \ \ x\to 0\ .
\eea
This solution can be analytically continued outside the domain
$\Re e\, \ell>(1+\alpha)\, L-{3/ 2}$. 
Let $\psi(x,E,-\ell-1)$ be another
solution of the same equation \eqref{Seq}
obtained from $\psi(x,E,\ell)$
by appropriate analytic continuation 
in $\ell$\ \footnote{In
general, analytic continuation 
of $\psi(x,E,\ell)$ in $\ell$ yields a
multivalued function of $\ell$, and $\psi(x,E,-\ell-1)$ at different
branches solves \eqref{Seq}\ 
with different potentials, corresponding to
different solutions of the same equations \eqref{algsys}.
However, since $\ell$ enters the 
Eq.\eqref{algsys}\ only through $\Delta$, Eq.\eqref{cdelta}\,,\ 
there exists a continuation $\ell\to -\ell-1$ 
which leaves
the potential\ \eqref{Vpot}\ unchanged.}.
For generic
values of $\ell$ the solutions $\psi_+(x,E)=\psi(x, E,\ell)$ and
$\psi_-(x,E)=\psi^+(x,E,-\ell-1)$
are linearly independent, since their Wronskian
$$W[\psi_{+}, \psi_{-}]=\psi_{+}\, \partial_x
\psi_{-}-\psi_{-}\, \partial_x \psi_{+}$$
equals to \ $-(2\ell +1)$.

For certain isolated values of $E$ one of
these solutions decays at $x\to +\infty$. Let
$\big\{E_k^{+}\big\}_{k=1}^{\infty}$ and 
$\big\{E_k^{-}\big\}_{k=1}^{\infty}$ be 
ordered spectral sets defined
by the conditions
\bea\label{eigenfun}
\psi_k^+(x)\equiv\psi^{+}(x,E^{+}_k)\to 0\, ,\qquad
\psi^{-}_k(x)\equiv\psi^{-}(x,E^{-}_k)\to 0\, ,
\eea
as\ $x\to+\infty$. 
One then defines
\bea\label{sdets}D_{\pm}(E)=\prod_{k=1}^{\infty}\Big(1-
{E\over E^{\pm}_k}\Big)\ .
\eea
Straightforward WKB analysis of the equation \eqref{Seq}\ shows that
\bea\label{eass}
E^{\pm}_k=\nu\ \left(k -{\textstyle \frac{1}{2}}\pm {\textstyle \frac{2l+1}{4} 
 }\right)^{2
\alpha\over 1+\alpha}\ \left(1+o(1)\right)\, ,
\ \ \ \  {\rm as}\ \ \ \
k\to\infty\ ,
\eea
and therefore at $\alpha > 1$ these products converge for all
finite $E$. For $0<\alpha\leq 1$ the definition\ \eqref{sdets}\ has
to be modified by including the Weierstrass factors, but
this does not affect our arguments.

The spectral determinant $D_{\pm}(E)$ thus defined are entire
functions of $E$. 
Asymptotic behavior of these functions at large $E$
follows from \eqref{eass},
\bea\label{Dass}
\log D_\pm(E)=
(-E/\nu)^{1+\alpha
\over 2\alpha}\mp {\textstyle \frac{2l+1}{4}}\, 
\log(-E)+O(1)\,,\quad
\arg(-E)<\pi\, .
\eea
And one can repeat the arguments of \cite{BLZf}\ to show that the
spectral determinants \eqref{sdets}\ satisfy 
exactly the same functional
relation  \eqref{wronsb},
as the direct consequence of the
properties $(I-III)$ of 
the Q-potential. We conclude that the pair of
functions
$D_{\pm}(E)$ satisfies, upon the 
identifications  \eqref{conj},
all the analytic conditions $(i-iii)$ of Section~2. 
This shows that every Q-potential
\eqref{Vpot}\ corresponds 
to some eigenvector of the ${\bf Q}$-operators, and
the associated eigenvalues $Q_{\pm}(s)$ coincide, up to the factors
$(-s)^{\pm {2\ell+1\over 4}}$, with the spectral
determinants $D_{\pm} (\nu s)$. 
Moreover,
using the WKB expansion of the solutions of 
\eqref{Seq}, it is possible to
evaluate corrections to the 
leading asymptotics \eqref{Dass}\ for any Q-potential
\eqref{Vpot}. For the
first such correction 
one obtains exactly \eqref{sublead}\ with $L$ standing for
the number of nonzero poles $z_1,\, z_2\,,
\ldots\,,z_L$ of \eqref{Vpot}. This confirms
the identification of $L$ in \eqref{Vpot}\ with the level.

At this point the question arises:
is this description of the spectrum
of the ${\bf Q}$-operators complete? In other words, is the family
of potentials \eqref{Vpot}\ 
large enough to accommodate all the eigenvalues
$Q_{\pm}(s)$? 
Answering this question requires enumerating all solutions
of the algebraic equations \eqref{algsys}\ 
for an arbitrary level $L$. We did
not accomplish this task. 
However, an elementary analysis of a few lowest
levels suggests positive answer to this question. 
For instance, for
$\Delta \neq 0$ the
level 1 subspace ${\cal V}_{\Delta}^{(1)}$ is
one-dimensional. Correspondingly, 
for $L=1$ the equation \eqref{algsys}\ has
only one solution
\bea\label{defs}
z_1 = z = {4\,(1+\alpha)\over\alpha}\ \Delta\, .
\eea
Therefore the spectral determinants 
$D_{\pm}(E)$ of \eqref{Seq}\ with
\bea\label{vone}
V(x)&=&x^{-2}\ 
\bigg\{\, x^{2\alpha+2}+\ell (\ell+1)+4 \alpha+4\nonumber\\&&\
\ \ \ \ \ \ \ \ \ \ \  
+{4 (\alpha+1)(2\alpha+3)\, z\over x^{2\alpha+2}-z}+
{8 (\alpha+1)^2 z^2\over (x^{2\alpha+2}-z)^2}\ \bigg\}
\eea
coincide with the eigenvalues of 
${\bf Q}_{\pm}(s)$ corresponding to
the eigenvector ${\bf L}_{-1}\, |\, \Delta\,  \rangle$. 
Note that $z$ turns
to zero at $\Delta =0$, 
in agreement with the fact that in this case
${\bf L}_{-1}\, |\, \Delta\, \rangle$ 
becomes a null-vector.

For $L=2$
the system \eqref{algsys}\  is 
solved by elementary methods.
For generic $c$
and $\Delta$ it has exactly two solutions (of course we do not
distinguish solutions which differ by permutations of
$z_k$).
We can recall that for
a non-degenerate Virasoro module the dimension of ${\cal
V}_{\Delta}^{(2)}$ equals 2. Again,  one of the solutions
disappears (in the sense that one or both of the numbers $z_1,\, z_2$
turn to zero) in three cases $\Delta = 0$, $\Delta={1-2\alpha\over
4(1+\alpha)}$ and
$\Delta
={1+3\alpha\over 4} $, as expected, because at these values of $\Delta$
null-vectors appear at the level 1 or 2, and the dimension of ${\cal
V}_{\Delta}^{(2)}$ drops down. Similar situation is observed at the level 3.

\section{Local integrals of motion}

Finally, let us make remark on the eigenvalues of the local
Integrals of Motion (IM) and their relation to
the semiclassical
analysis of the \SE\ with the Q-potentials \eqref{Vpot}.
Asymptotic expansions of the operators ${\bf Q}_{\pm}(s)$
around the essential singularity at $s=\infty$ (in the sector
$-\pi < \arg(-s) < \pi$) are expressed in terms of the infinite
sequence of the local IM ${\bf I}_{2n-1}$ and
``dual nonlocal IM'' $\widetilde{{\bf H}}_{n}^{\pm}$ \cite{BLZb},
\bea\label{asympt}
\log {\bf Q}_\pm(s)&\simeq&
{\pi\over \cos({\pi\over 2\alpha})}\
 (-s)^{\frac{1+\alpha}{2\, \alpha}}
+ {\bf C}+
\sum_{n=1}^{\infty}\, 
(-\nu s)^{\frac{(1-2 n)(1+\alpha)}{2\, \alpha}}\ b_{n}\
 {\bf I}_{2n-1}\nonumber\\&&+
\sum_{n=1}^{\infty}
\,
(-\nu s)^{-n(1+\alpha)}\  c_n \
\widetilde{{\bf H}}^\pm_{n} \ ,
\eea
where the coefficients $b_n$ and $c_n$ have explicit form
\bea\label{Bdef}
b_n&=&{(-4)^{n-1}\over n!\alpha \sqrt{\pi}}\ 
(1+\alpha)^n\  
\Gamma\Big({1-2n\over 2\alpha}\Big)
\Gamma\Big({(2n-1)(1+\alpha)\over 2\alpha}\Big)\, ,\nonumber
\\
c_n&=&(-1)^{n-1} \ 4^{n\alpha}
\ (1+\alpha)^{-2n}
\ .\eea
The asymptotic expansions of the eigenvalues $Q_{\pm}(s)$ are thus
expressed through the eigenvalues of the local and nonlocal IM.
Note that \eqref{asympt}\ has exactly the structure which appears
when one performs the WKB expansion of the solutions of the \SE\
\eqref{Seq}\ with a Q-potential \eqref{Vpot}.
The correspondence \eqref{conj}
suggested in the previous section implies that the coefficients
generated by this WKB expansion must agree with the eigenvalues of the
operators ${\bf I}_{2n-1}$ and $\widetilde{{\bf H}}_{n}^{\pm}$.

Whereas the eigenvalues of the dual nonlocal IM are relatively
difficult (but not impossible) to calculate, the local IM ${\bf
I}_{2n-1}$ have explicit expressions
in terms of the Virasoro generators ${\bf L}_n$ (see Eq.(11) of
Ref.\cite{BLZa}),
and their eigenvalues in any given level subspace
${\cal V}_{\Delta}^{(L)}$ can be found by a direct calculation.
Thus, for the vacuum
eigenvalues of the first few ${\bf I}_{2n-1}$ we have \cite{BLZa}
\bea\label{vacint}
I^{(vac)}_1(\Delta)&=&\Delta-{c\over 24}\, ,\nonumber\\
I^{(vac)}_3(\Delta)&=&\Delta^2- {c+2\over 12}\ \Delta
+{c\, (5 c+22)\over 2880}\, ,\\
I^{(vac)}_5(\Delta)&=&
\Delta^3-{c+4\over 8 }\ \Delta^2+
{(c+2) (3 c+20)\over 576}\
\Delta  \nonumber\\ &&\,\,-\,\,\,{c (3 c+14) (7 c+68)\over 290304
}\, ,\nonumber
\eea
where the Virasoro central charge
$c$ is related to the parameter $\alpha$ as in Eq.\eqref{cdelta}.
We have checked by explicit diagonalization at the levels $L=1,2,3$
that the eigenvalues of the first three local IM agree with the
following expressions
\bea\label{sumrules}
I_1&=&I_1^{(vac)}(\Delta+L)\, ,\nonumber
\\
I_3&=&I_3^{(vac)}(\Delta+ L)+{\alpha\over
1+\alpha}\ \sum_{k=1}^L\, z_k\ ,\\
I_5&=&I_5^{(vac)}(\Delta+L)+
\Big(\,  {\Delta+L-1\over 5+2\alpha}+{3+\alpha
\over 24(1+\alpha)}\,
\Big)\times \nonumber
\\ &&
{3\alpha (3+2\alpha)\over 1+\alpha}\ \sum_{k=1}^L\, z_k
+{3\, \alpha^2\over
2(5+2\alpha)(1+\alpha)^2}\  \sum_{k=1}^L\, z^2_k\ ,\nonumber
\eea
in terms of the same parameters $z_k$ that enter \eqref{Vpot}.
One can verify that the same expressions come out from
the WKB analysis of \SE\ \eqref{Seq},\eqref{Vpot}.

\bigskip
\section*{Acknowledgments}

\noindent
The authors thank C.~Ahn for reading the manuscript and important remarks.
VVB thanks the Department of Physics and Astronomy at Rutgers
University for the hospitality
extended to him during his visit to Rutgers in April 1999 when
most of the results of this paper have been obtained. He also thanks
A.N.~Kirillov for interesting discussions. 

\bigskip

\noindent
The research of SLL and ABZ is supported
in part by DOE grant $\#$DE-FG02-96 ER 40949.

\bigskip
\bigskip

\renewcommand{\theequation}{\thesection.\arabic{equation}}
\renewcommand{\thesection}{\Alph{section}}

\section*{Appendix A}
\setcounter{section}{1}
\setcounter{equation}{0}

In this Appendix we assume that $\alpha$ and $\ell$ take some generic
values so that the Virasoro module ${\cal V}_\Delta$ is
non-degenerate. Moreover we assume that $\alpha>1$
\footnote{ For $0<\alpha\le1$
the definition\ \eqref{logdef},\eqref{phidef}
below may need modification, but the statement about the one to one
correspondence between the eigenstates and the
sets \eqref{mpk},\eqref{Ndef}  most likely remains valid.}.
Then the
entire functions ${A}_{\pm}(s)=(-s)^{\mp {2\ell+1\over 4}}
\,{Q}_{\pm}(s)$ are
uniquely determined
by their leading asymptotics at $s\to \infty$
\eqref{aslead}, the normalization\ \eqref{normalki}, and the locations of
their respective zeroes $s_k,\ (k=1,\, 2,\ldots)$ in the complex $s$-plane.

From the quantum Wronskian relation \eqref{wronsb}\ it 
follows that for any eigenvector of ${\bf Q}_{\pm}(s)$  the zeroes
${s_k^{\pm}}$ of
${A}_{\pm}(s)$
satisfy the same infinite system of the  Bethe Ansatz equations
\bea\label{BAeq}
{{A}_{\pm}(s^\pm_k\,q^2\ )\over{A}_{\pm}(s^\pm_k\,q^{-2})}=-
q^{\mp(2\ell+1)}\ .
\eea
The eigenvalues
corresponding
to  different eigenvectors
are distinguished by a specific phase assignment in this
equation
\bea\label{BAlog}
{1\over 2\pi {\rm i}}\
\log\left[\, {{A}_{\pm}(s^\pm_k\,q^2\
)\over{A}_{\pm}(s^\pm_k\,q^{-2})}\, \right]
=\mp {2\ell+1\over 2(1+\alpha)}-m^\pm_k
+{1\over2}\ ,
\eea
determined by a choice of integers  $\{m^\pm_k\}_{k=1}^\infty$.
These integers, of course, depend on the choice of branches of the
logarithm 
in the left hand side
of \eqref{BAlog}. However, once these branches are appropriately
fixed (see below),
every eigenvalue  is characterized by a unique
set of the integers $\{m^\pm_k\}$.

As follows from $(ii)$ in the Section 2, the zeroes $s^{\pm}_k$
accumulate toward infinity along the positive real axis. In fact, explicit
solutions for special values of $c$, as
well as numerical analysis, give all support to the following

\bigskip
{\bf Conjecture}:
For all eigenvalues at the level $L$, and for
real $\ell>(1+\alpha)L-3/2$ all zeroes $s^{+}_k$ 
of   $ A_{+}(s)$ are simple, real and positive.
The zeroes of $ A_{-}(s)$ enjoy the same property for
$\ell< -(1+\alpha)L+3/2$.

\bigskip
		
Assuming this property one can 
unambiguously fix the phase of the logarithm in \eqref{BAlog}.
Consider, for example, the zeroes $\{s^+_k\}$ of $A_+(s)$
for $\ell>(1+\alpha)L-3/2$ and define
the L.H.S. of Eq.\eqref{BAlog} as
\bea\label{logdef}
{\rm L.H.S.\ \ of\ \ Eq.\eqref{BAlog}}
={1\over \pi}\ \sum_{j=1}^\infty\arg(1-q^2 s_k/s_j)\, ,
\eea
where for real positive $x$
\bea\label{phidef}
-\pi<\arg(1-q^2 x)<+\pi\, .
\eea
Since all zeroes $s_k^+$ are assumed to be
distinct and positive  they can be uniquely ordered by their value
$s^+_1<s^+_2<s^+_3<\ldots$\ . Then for each eigenvalue $A_+(s)$ 
all the  integers $m_k^+$ will be distinct and uniquely defined. 
For other values of $\ell$ 
the L.H.S. of \eqref{BAlog}\ is defined by means of
the analytic continuation of \eqref{logdef}\ in $\ell$.

Obviously, not every set of integers
$\{m^\pm_k\}$ corresponds to an eigenstate of
${\bf Q}_{\pm}(s)$ in ${\cal V}_\Delta$. Indeed, 
it follows from\ \eqref{aslead}\ that
\bea\label{zerasympt}
s^\pm_k=\left(k -{\textstyle \frac{1}{2}}\pm {\textstyle \frac{2l+1}{4}
 }\right)^{2
\alpha\over 1+\alpha}\ \left(1+o(1)\right)\, ,\qquad
{\rm as}\qquad
k\to\infty\, .
\eea
Hence\ \eqref{BAlog} implies that for a given
eigenvalue the sequence of  integers $m_k^{\pm}$ stabilize at
large $k$, i.e. $m_k^{\pm}=k$ 
for sufficiently large $k$.
The sets $\{m_k^{\pm}\}$ associated with different eigenvalues differ in
finitely many first entries.
Therefore the most general pattern in  the set 
 $\{m_k^+\}$ (or  $\{m_k^-\}$) can be obtained from the vacuum  set 
($m_k^+= k$, for all $k=1,2,\ldots,\infty$) by deleting 
a certain number of entries (we denote this number by $M$) 
and adding the same number
of distinct non-positive integer entries. It can be written as
\bea\label{mpk}
m^+_k=\left\{
\begin{matrix}
1-n^-_{M-k+1}\ , &{\rm for\ \ }
k=1,\, \ldots M\ ,\hfill\\
N_{k-M}(n^+)\ ,&{\rm for\ \ } k\ge M+1\ .\hfill
\end{matrix}
\right.
\eea
Here $n^\pm=\{n^\pm_1,n^\pm_2,\ldots,n^\pm_M\}$ denote two increasing
sequences  of positive integers
$1\le n^+_1<n^+_2<\ldots<n^+_M$ and $1\le n^-_1<n^-_2<\ldots<n^-_M$, 
with $M\ge0$;\   
and $N_j(n^+)$,\  $j=1,2,\ldots,\infty$, \   denotes $j$-th
element of the increasing sequence of
consecutive positive
integers with deleted entries $n^+_i$, $i=1,\ldots M$:
\def\crossout#1{%
\hbox{\big/}\kern-.73em {#1}}
\bea\label{Ndef}
\{N(n^+)\}=\{\, 1,\, 2,\, \ldots\, \crossout{n^+_1},\, \ldots\ ,
\crossout{n^+_2},\, \ldots\, \}\ .
\eea
Using the standard technique
based on the Destri-de~Vega 
equation \cite{DdV95,KBP} 
associated with \eqref{BAlog} and \eqref{mpk} one can easily
calculate the next  
subleading term in the large $s$ asymptotics of the corresponding eigenvalues 
$A_+(s)$. The result is given by \eqref{sublead} 
with
\bea\label{npm}
L=\sum_{i=1}^M\, (n^+_i+n^-_i-1)\ . 
\eea
For a given value of $L$ there  are exactly $\prt(L)$
sets ${\cal S}^{(L)}_k=\{n^+\vert n^-\}$, with  $k=1, 2, \ldots
\prt(L)$,  satisfying \eqref{npm}\footnote{
The counting follows
from the following elementary partition identity
$$\sum_{n=0}^\infty \prt(n)\  z^n =\sum_{n=0}^\infty
{z^n\over(z)_n}
 =\sum_{n=0}^\infty {z^{n^2}\over(z)_n^2}\, , \qquad
(z)_n\equiv \prod_{k=1}^n
(1-z^k)\, \qquad (\, |z|<1)\ .
$$}, where $\prt(L)$ stands for the number of integer partitions of $L$.
These sets naturally enumerate all eigenvectors 
in the
level subspace ${\cal V}^{(L)}_\Delta$\ \footnote{
For example, at $L=4$, there are five sets ${\cal S}^{(4)}_k$,\ 
$\{4\vert 1\}, \{1\vert 4\}, \{2\vert 3\}, \{3\vert 2\},
\{1,2\vert 1,2\}$.}.
As for the set $\{m^-_k\}$ describing the zeroes of $ A_-(s)$,
it is obtained from $\{m^+_k\}$ by interchanging the roles of 
$n^+$ and $n^-$ in  Eq.\eqref{mpk}.

As explained in \cite{BLZf}\ the formula  \eqref{asympt}\ allows 
one to derive the asymptotic expansions for the zeroes
of the eigenvalues ${A}_\pm(s)$. Indeed, let $I_{2n-1}$ and
$\widetilde{H}^\pm_n$ are the
eigenvalues of the local
IM  and the dual nonlocal IM  corresponding to a
certain eigenvector of ${\bf Q}_\pm(s)$ determined by the set of
phases \eqref{mpk}. Then substituting \eqref{asympt}\ into
\eqref{BAlog}\ one obtains
\bea\label{zeroas}
s_k^{\frac{1+\alpha}{2\, \alpha}}+
\sum_{n=1}^\infty\,
(\nu s_k)^{\frac{(1-2 n)(1+\alpha)}{2\, \alpha}}\ \beta_{n}\
 { I}_{2n-1}+ 
\sum_{n=1}^{\infty}
\,
(\nu s_k)^{-n(1+\alpha)}\  \gamma_n \
\widetilde{{ H}}^\pm_{n}\,\ \ \ \ \ &&\nonumber  \\
\simeq\, 
m_k^{\pm} -{\textstyle \frac{1}{2}}\pm {\textstyle \frac{2l+1}{4}}\,,&& 
 \eea
where $k=1,2,\ldots\,$ and 
\bea\label{bgdef}
\beta_n&=&(-1)^n\  \ \pi^{-1}\ \cos\big({\pi (n-1/2\big)/ 
\alpha})
\ \ b_n\, ,\nonumber\\ 
\quad \gamma_n&=& (-1)^{n+1} \ \pi^{-1}\  \sin(\pi n\alpha)
\ \ c_n\, .
\eea
On the other hand the Bethe Ansatz
equations \eqref{BAlog}\ complemented by the asymptotic condition 
\eqref{zerasympt} can be solved numerically \cite{Voros-b}.
We have performed these calculations for the zeroes of
$A_+(s)$ in a particular case
$\alpha=4$ with various choices of the parameter $\ell$ and
the integer phases $\{m^+_k\}$ as in\ \eqref{mpk}\ above.
For each of these choices we then
calculated  a few first  coefficients in the asymptotic expansion
\eqref{zeroas}\ and thus determined the eigenvalues $I_{2n-1}$ numerically
(Note, that for $\alpha=4$ the second sum in \eqref{zeroas}\ vanishes.). The
results are in very good agreement with those obtained from an explicit
diagonalization of the first three local IM \eqref{sumrules}. The
correspondence was verified  for all eigenvectors
at the level $L\le5$.

\section*{Appendix B}
\setcounter{section}{2}
\setcounter{equation}{0}

Here we explain why the form \eqref{Vpot}\ describes the most general
potential which satisfies the conditions $(I-III)$ in  Section 3.
Indeed, let us write $V(x)$ in the form
\bea\label{newpot}
V(x)={\ell(\ell+1)\over x^2}+x^{2\alpha}+v (x)\ .
\eea
The conditions $(I-III)$ imply that
\bea\label{vform}
v(x)=x^{-2}\, F\big(x^{2\alpha+2}\big)\ ,
\eea
where $F(x^{2\alpha+2})$ is a rational function of its argument,
vanishing at $x=0$ and bounded at $x=\infty$,
\bea\label{fcond}
F(0)=0\, ,\qquad |F(\infty)|<\infty\ .
\eea
Therefore, unless
$F\equiv 0$, the potential \eqref{newpot}\ has poles at finite nonzero values
of $x$.  Let $x_0\neq 0$ be the position of any such pole.
Consider the Laurent expansion of the potential \eqref{newpot}\ near 
this pole
\bea\label{laur}
V(x)=\sum_{k=r}^\infty \,(x-x_0)^k\ V_k\ .
\eea
It is well known (see e.g. \cite{DG}) that coefficients of this expansion
should be constrained as
\bea\label{ccond}
r\ge-2\, ; \qquad V_{-2}=m(m+1)\, ;\qquad
V_{2k-1}=0\,, \qquad k=0,\, 1,\ldots m\ ,
\eea
where $m=0,\,  1,\, 2\ldots$,
to ensure that both linear independent
solution of \eqref{Seq}\ are single valued functions  $x$ in the vicinity of
$x_0$ for all values of $E$. One can check that for $m\ge2$ there are
too many conditions in \eqref{ccond}\ which cannot be simultaneously
satisfied for generic values of
$\alpha$ and $\ell$. So the only case to consider is $m=1$ which
requires
\bea\label{onlycase}
V_{-2}=2\, ,\ \qquad V_{-1}=V_{1}=0\, ,
\eea
for every pole of \eqref{newpot}.
These conditions, together with \eqref{vform}\ and
\eqref{fcond}, lead to  \eqref{Vpot}\ and \eqref{algsys}.

\end{document}